\documentclass[12pt]{article}
 
 \usepackage{graphicx}
 \usepackage[utf8]{inputenc}
 \usepackage{psfrag}
 \usepackage{amsmath,amssymb,amsfonts,amsthm}
 \usepackage{booktabs}
 \usepackage{url}
 \usepackage{lscape}
 \usepackage{setspace}
  \usepackage{multirow}

\usepackage[normalem]{ulem}

 \usepackage{authblk}
 \usepackage{empheq} 
 \usepackage{bbm} 

 \usepackage[round,sort]{natbib}

 \def \IR{\hbox{{\rm I}\kern-.2em\hbox{{\rm R}}}}
 \newcommand{\mv}[1]{{\boldsymbol{\mathrm{#1}}}}

 \usepackage{amsthm}  
 \usepackage{algorithm}
 \usepackage{algpseudocode}

\usepackage{moreverb}

\newcommand\BibTeX{{\rmfamily B\kern-.05em \textsc{i\kern-.025em b}\kern-.08em
T\kern-.1667em\lower.7ex\hbox{E}\kern-.125emX}}

\usepackage{cases}
\usepackage{ulem}

 \providecommand{\keywords}[1]{\textbf{Keywords:} #1}
 
 \graphicspath{{figs/}{./}}
 \theoremstyle{remark}
 
 \usepackage{color}
\title{Dynamic predictions of kidney graft survival   
in the presence of longitudinal outliers}
\author[1*]{Özgür Asar}
\author[2]{Marie-Cécile Fournier}
\author[2]{Etienne Dantan}
\affil[1]{Department of Biostatistics and Medical Informatics, 
Ac{\i}badem Mehmet Ali Ayd{\i}nlar University, \.{I}stanbul, Turkey.}
\affil[2]{INSERM UMR 1246 - SPHERE, Nantes University, Tours University, Nantes, France.}

\date{* ozgur.asar@acibadem.edu.tr $|$ ozgurasarstat@gmail.com}

\begin{document}

\maketitle
  
\begin{abstract}
 
\noindent In kidney transplantation, dynamic predictions of graft survival may be obtained from joint modelling of longitudinal and survival data for which a common assumption is that random-effects and error terms in the longitudinal sub-model are Gaussian. However, this assumption may be too restrictive, e.g. in the presence of outliers, and more flexible distributions would be required. 
In this study, we relax the Gaussian assumption by defining a robust joint model with $t$-distributed random-effects and error terms to get dynamic predictions of graft survival for the kidney transplant patients from the French DIVAT cohort. We take a Bayesian paradigm for inference and dynamic predictions and sample from the posterior densities. While previous research reported improved performances of robust joint models compared to the Gaussian version in terms of parameter estimation, dynamic prediction accuracy obtained from such approach has not been yet evaluated. Our results illustrate that estimates for the slope parameters in the longitudinal and survival sub-models are sensitive to the distributional assumptions. From a validation sample, calibration and discrimination performances appeared better under the robust joint models compared to the Gaussian version, illustrating the need to accommodate outliers in the dynamic prediction context.

\end{abstract}

 
 \section{Introduction}
 \label{sec:introduction}
 
In the context of chronic diseases, prediction scores of clinical events have become increasingly popular. These scores may help patients and physicians in a shared decision-making and facilitate the implementation of the P4-medicine (predictive, personalized, preventive and participative) \citep{Flores2013} in clinical practice. Often measured to assess patients' health evolution during the follow-up, longitudinal markers can be used to improve time-fixed (static) predictions obtained using only baseline information. Dynamic predictions are therefore defined as updated predictions, whenever any new data become available along the follow-up \citep{rizopoulos_joint_2012, proust-lima_dynamic_2014}. 

 
In the statistical literature, there is a growing interest in methods to compute dynamic predictions. Among them, joint modelling of longitudinal and survival data is one of the most popular \citep{proust-lima_development_2009, rizopoulos_dynamic_2011, taylor2013, asar2015}. Classically, survival outcomes are modelled using a Cox model with a time-varying frailty term, whereas the repeated measures are modelled using a mixed-effect model with Normally distributed random-effects and error terms. In longitudinal clinical  studies, some observations may be highly apart from the others, and two types of outliers may be defined: i) at population level, outlying subjects who do not follow the typical population trend, ii) at individual level, outlying observations that do not follow the typical trajectory of an individual \citep{pinheiro_efficient_2001, wu_mixed_2009}. Gaussian assumption would not give appropriate weights to these individuals or observations, i.e. it would not be robust against outliers \citep{sutradhar_estimation_1986, lange_robust_1989}. 
  
In the context of kidney transplantation, a systematic review on predictive models emphasized the need for dynamic predictions  \citep{Kabore2017}. Since longitudinal measures of serum creatinine (SCr) have been demonstrated as associated with kidney graft failure \citep{Fournier2016c}, we recently proposed to use them to obtain dynamic predictions of long-term kidney graft failure \citep{fournier_ndt}. For the French kidney transplant cohort DIVAT (www.divat.fr), we obtained 
dynamic predictions of graft survival based on a joint model with Gaussian random-effects and error terms.  Nevertheless, such dynamic predictions based on this joint
model may be sub-optimal in presence of longitudinal outliers.

The objective of the current paper is to investigate possible impacts of longitudinal outliers on dynamic predictions of long-term kidney graft survival. We compare prognostic accuracies (discrimination and calibration) of the Gaussian and robust joint modelling approaches for the DIVAT patients. For the robust joint modelling, we relax the Gaussian assumption by postulating $t$ distributions for the random-effects and error terms. The robust model we consider is novel in the sense that it postulates independent $t$ distributions for the random-effects and error terms, and error terms are allowed to be independent within a subject. To the best of our knowledge, there is no work in the literature that considered such a joint model. For inference and dynamic predictions, we take a Bayesian paradigm and sample from the joint posterior densities. While several authors reported better performance of robust joint models compared to the Gaussian version in terms of parameter estimation, especially in terms of standard error estimation \citep{li_robust_2009, huang2010, kim_sungduk2016, baghfalaki2013, baghfalaki2014}, the advantages of robust joint modelling regarding individual dynamic predictions are yet to be explored. 
 
 
The rest of the paper is organised as follows. 
In Section \ref{sec:data}, we give details of the DIVAT data 
that motivates this work. 
Sections \ref{sec:joint_model} 
introduces the general modelling framework, distributional 
assumptions and inferential procedures. 
Section \ref{sec:dynamic_prediction} describes 
dynamic predictions for newcomer subjects and 
accuracy measures to evaluate prognostic capabilities 
of the models. 
In Section \ref{sec:application}, we present 
a concrete application to the DIVAT data-set. 
Section \ref{sec:discussion} closes the paper. 



 \section{DIVAT data}
 \label{sec:data}
 
Data were extracted from the multicentric French kidney transplant cohort DIVAT (French Research Ministry: RC12\_0452, last agreement No 13 334, No CNIL for the cohort: 891735). The inclusion criteria are: adult recipients who received a first or second renal graft from a living or heart-beating deceased donor, alive with a functioning graft at 1 year post-transplantation, and maintained under Tacrolimus and Mycofenolate. The aim is to predict graft failure risk in the chronic phase of transplantation. Therefore, time origin is selected as 1 year after transplantation. Graft failure is defined as the occurrence of any of the following events: return to dialysis, pre-emptive re-transplantation, or death with a functioning graft. Serum creatinine (SCr, measured in $\mu$mol/L) is the longitudinal biomarker, 
with high values indicating worse kidney health.  
Following  \citet{fournier_ndt}, the following set baseline covariates are considered: recipient age (Age: in years), history of cardiovascular disease (CV: yes/no), 3-month SCr (SCr3), occurrence of an acute rejection episode in the first year post-transplantation (AR: yes/no), pre-transplantation anti-class I immunisation (ACI: yes/no) and graft rank (GR: second/first). Follow-up time $(t_{ij})$ is the time-varying covariate. 
 
\begin{figure}[t]
\begin{center}
\includegraphics[scale = 0.8]{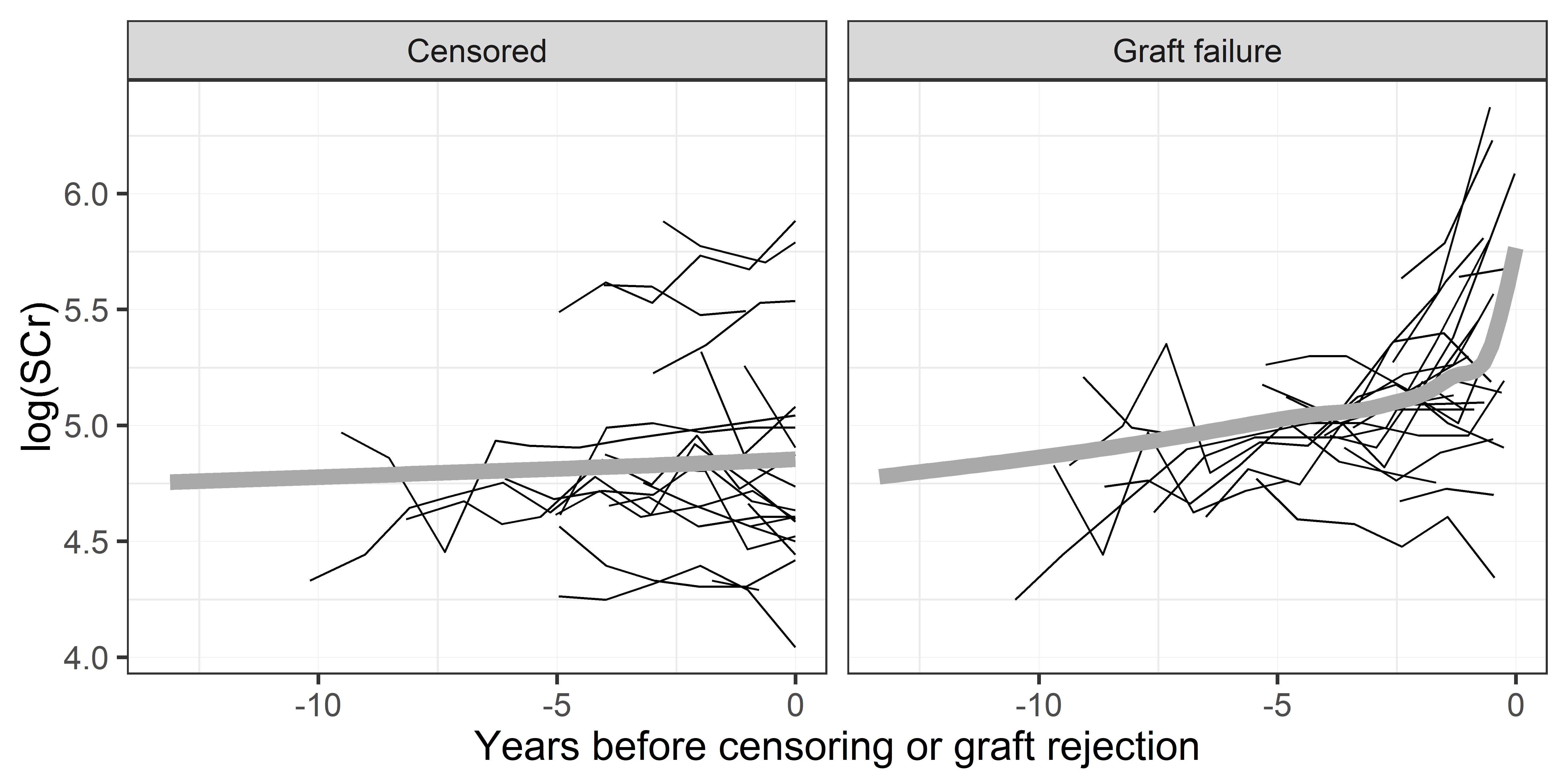}
\caption{Spagetti-plots for the learning sample. 
x-axis is aligned time obtained 
as follow-up time minus 
censoring or graft failure time. 
Solid black lines are individual trajectories for randomly 
selected subjects; 30 patients in the censored group and 30 in the 
graft failure group.  
Super-imposed thick gray lines are fitted LOWESS curves to the 
whole data in each group; 2,127 patients in censored group, 457 in the graft failure.  
}
\label{fig:spagetti_train_subset}
\end{center}
\end{figure}

\begin{figure}[t]
\begin{center}
\includegraphics[scale = 0.85]{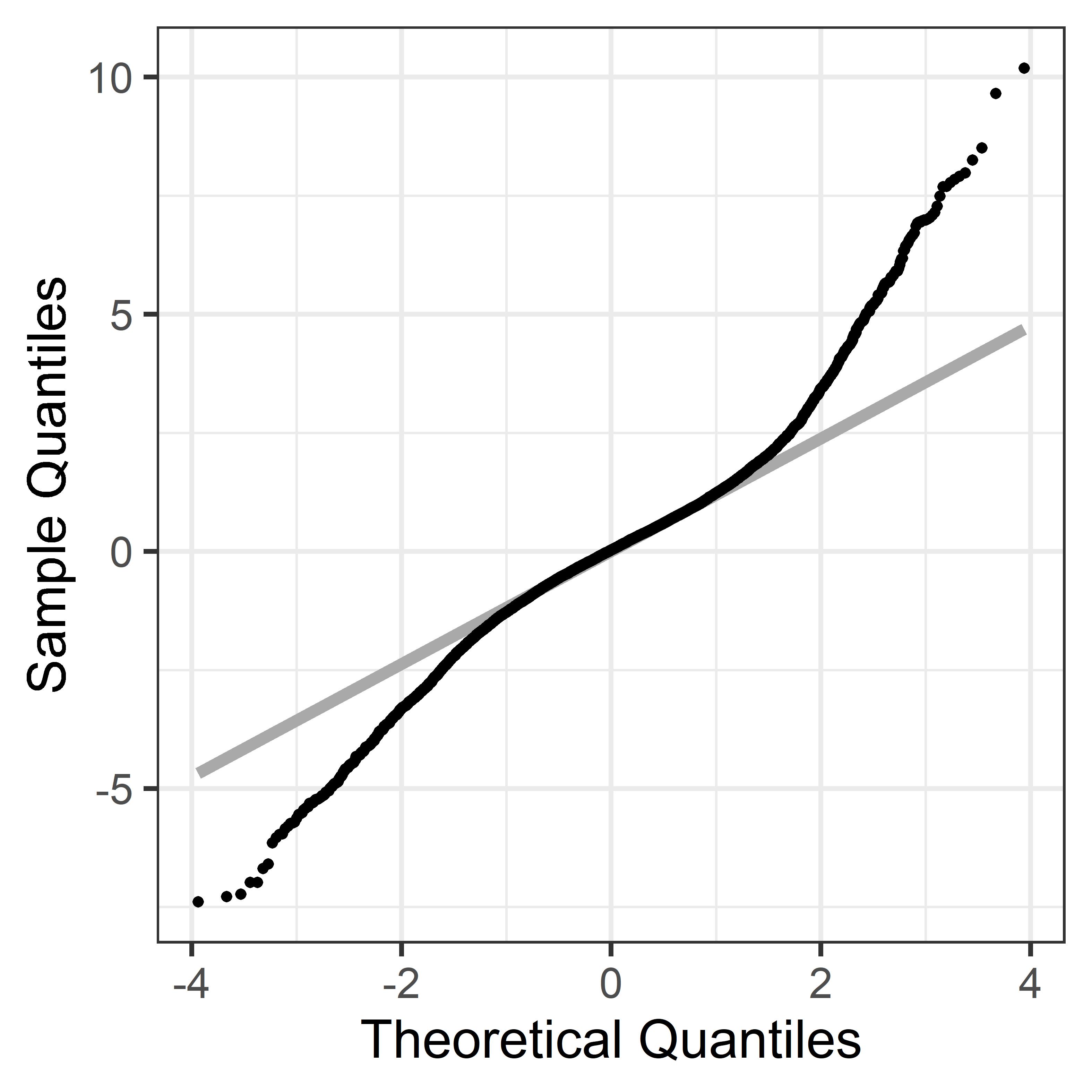}
\caption{Quantile-quantile plot of the standardised 
marginal residuals (based on the 
joint model with Normal distributional assumptions 
for the random-effects and error terms) 
against the standard Normal distribution.}
\label{fig:train_qq}
\end{center}
\end{figure}

The learning and validation samples are constructed based on two data extracts. While the first extract covers the period of 1 January 2000 -- 31 August 2013, the second covers 1 September 2013 -- 31 October 2016. Randomly chosen two-third of the first extract constitutes the learning sample (2,584 patients), whereas remaining portion of the first extract plus the whole data from the second extract constitute the validation sample (1,577+946 = 2,523 patients). Number of repeated measures ranged between 1 and 14 (with median of 4) for the learning sample. Spagetti-plots SCr (in log-scale) for a random sample of 60  patients from the learning sample are shown in Figure \ref{fig:spagetti_train_subset}. For more details of the data, reader is referred to \citet{fournier_ndt}. 
The quantile-quantile plot displayed 
in Figure \ref{fig:train_qq} 
 indicates that there are considerable departures from the Gaussian assumption for the DIVAT data-set. 
   
\section{Joint modelling of longitudinal and survival outcomes}
\label{sec:joint_model}

 From Figure \ref{fig:train_qq}, 
 we are not able to discern the source of 
 heavy-tailedness,   
 i.e. whether it is due to 
 outlying individuals or outlying observations or both. 
 Therefore, in what follows we consider 
 a general modelling 
 framework which postulates that both types of 
 outliers may be present. 
 Based on the posterior summaries 
 for the degrees-of-freedom parameters 
 for random-effects and error terms,  
 we can then drop the heavy-tailedness assumption 
 for the term for which Normality is indicated.  
 
\subsection{Notations and framework} 

To make inference, we assume to observe a sample of $n$ independent and identically distributed subjects with the data of $\left\{T_i, E_i, Y_i, t_i, a_i; i = 1, \ldots, n \right\}$. Here, $Y_i = \left\{Y_{ij}; j = 1, \ldots, m_i \right\}$ is the set of longitudinal marker with the set of corresponding timings, $t_i = \left\{t_{ij}; j = 1, \ldots, m_i \right\}$; $a_i = \left\{a_{ik}; k = 1, \ldots, f \right\}$ the baseline explanatory variables; and $T_i$ the time elapsed between the origin and occurrence of the survival event. $T_i$ is typically right-censored, i.e. $T_i = \min(T_i^*, C_i)$, where $T_i^*$ being the true survival time and $C_i$ the censoring time for subject $i$. Therefore, an event indicator, $E_i = \mathbbm{1}\{T_i^* \leq C_i\}$, where $\mathbbm{1}\{\cdot\}$ denotes the indicator function, completes the survival information.  

In this study, we consider the so-called shared-parameter version of the joint model \cite{wulfsohn1997}, based on two linked sub-models.  The modelling framework can be defined as follows: 
\begin{numcases}{}
\label{eq:eq_long}
Y_{ij} = Y_i^{*}(t_{ij}) + Z_{ij} = {\mv x}_{ij}^{\top} {\mv \alpha} + {\mv d}_{ij}^{\top} {\mv B}_i + Z_{ij},&\\
\label{eq:eq_surv}
 h_i(t)= h_0(t) \exp({\mv c}_i^{\top} {\mv \omega} + 
 g\left(\mathcal{Y}_i^{*}(t), {\mv \eta}\right)).
\end{numcases}
Equation \eqref{eq:eq_long} corresponds to a linear mixed-effects model that defines the longitudinal process. The observed longitudinal measure, $Y_{ij}$, is assumed to be a noisy version of the underlying continuous-time signal, $Y_{i}^{*}(t)$, at time 
$t_{ij}$, with $Z_{ij}$ being the noise, or as often called measurement error. ${\mv x}_{ij} = [x_{ij1} \ldots x_{ijp}]^{\top}$ is a design matrix that consists of elements from $a_i$ and $t_i$, and ${\mv d}_{ij} = [d_{ij1} \ldots d_{ijq}]^{\top}$ another design matrix that is typically structured as a subset of ${\mv x}_{ij}$. ${\mv \alpha} = [\alpha_1 \ldots \alpha_p]^{\top}$ and ${\mv B}_i = [B_{i1} \ldots B_{iq}]^{\top}$ are population-averaged parameters and subject-specific random-effects (latent variables), respectively. 

Equation \eqref{eq:eq_surv} corresponds to a Cox model with time-varying frailty for the time-to-event. $h_i(t)$ is the instantaneous risk of experiencing the event at time $t$, and $h_0(t)$ the baseline hazard. $h_0(t)$ can be left un-specified as in \citet{cox1972}, or specified parametrically using hazard function of a life-time distribution, e.g. for Weibull $h_0(t) = \lambda \nu t^{(\nu - 1)}$, or using piecewise-constants or splines; for details see \citet{rizopoulos_joint_2012}. ${\mv c}_i = [c_{i1} \ldots c_{ir}]^{\top}$ is a design matrix that is composed of elements from $a_i$, and ${\mv \omega}$ are the associated parameters. $g\left(\mathcal{Y}_i^{*}(t), {\mv \eta}\right)$ is the term that links features of underlying signal up to and including time $t$, $\mathcal{Y}_i^{*}(t)$, and hazard function at time $t$ through a known link function $g(\cdot)$ and parameters ${\mv \eta}$. There are a number of choices for $g\left(\mathcal{Y}_i^{*}(t), {\mv \eta}\right)$. Widely used examples include, among others, the current value parametrization, $g\left(\mathcal{Y}_i^{*}(t), {\mv \eta}\right) = \eta Y_i^{*}(t)$, or current value and rate of change parametrization, $g\left(\mathcal{Y}_i^{*}(t), {\mv \eta}\right) = \eta_1 Y_i^{*}(t) + \eta_2 \frac{\partial Y_i^{*}(t)}{\partial t}$; for other parametrizations, see \citep{rizopoulos_joint_2012}. 

\subsection{Distributional assumptions}
\label{sec:distributional_assumptions}

Standard joint models assume that  
${\mv B}_i$ and $Z_{ij}$ are both zero-mean 
Gaussian, such that 
${\mv B_i} | {\mv \Sigma} \sim \mathcal{MVN} \left({\mv 0}, {\mv \Sigma} \right)$ 
and $Z_{ij}|\sigma \sim \mathcal{N}(0, \sigma^2)$.  
The terms are further assumed to have the following properties, 
${\mv B}_i \perp Z_{ij}$ and $Z_{ij} \perp Z_{ij'}$ 
for $j \neq j'$. Gaussian assumption might be too  
restrictive for some real-life applications,  
because the data-sets typically consist of subjects 
that exhibit outlying behaviours. 

\begin{figure}[t]
\begin{center}
\includegraphics[scale = 0.65]{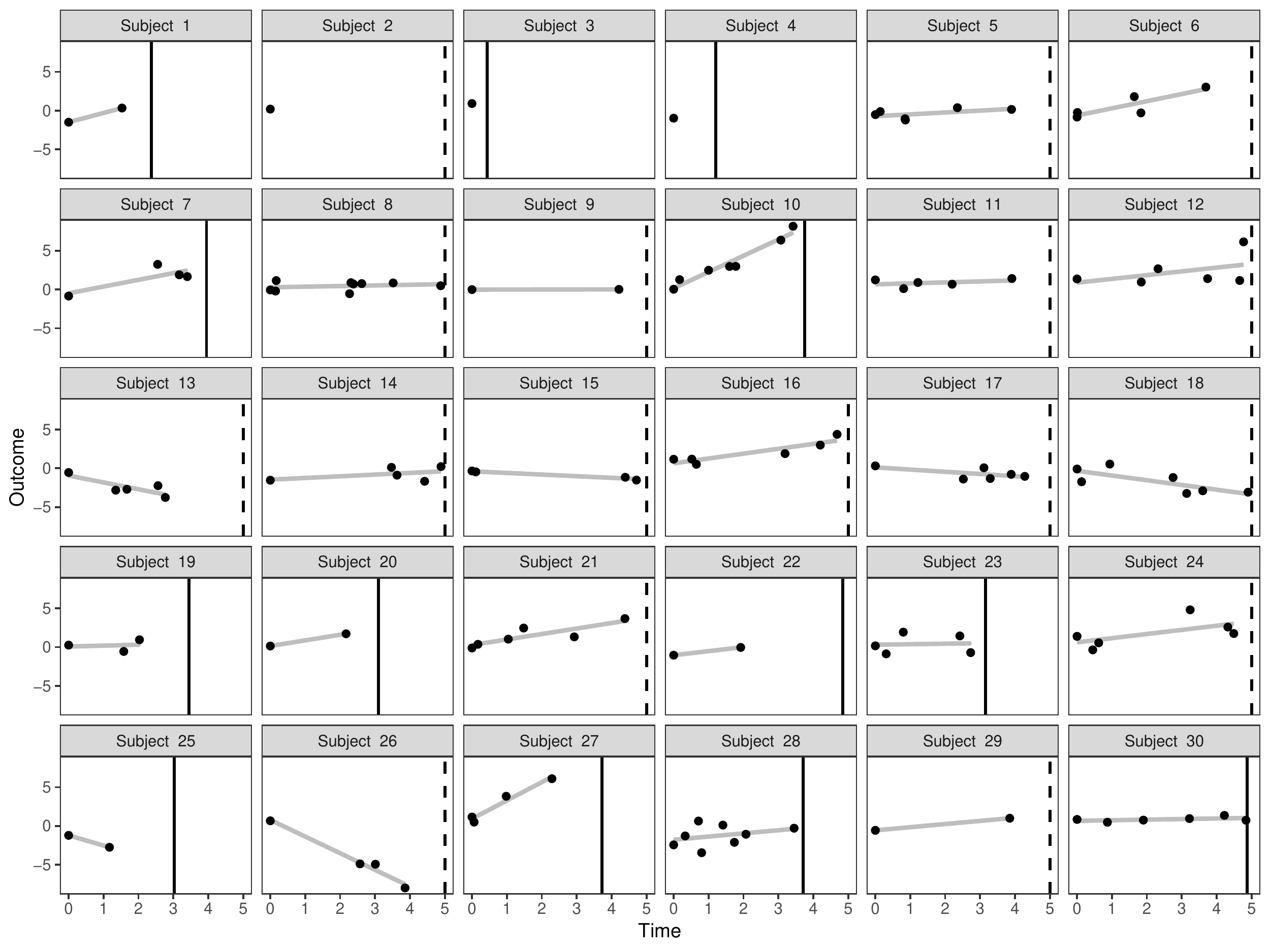}
\caption{Simulated data for 30 subjects, from a robust joint model with the population-averaged intercept and 
slope of 0. 
Dots are repeated measures. Dashed vertical lines indicate survival times for censored subjects, whereas solid vertical lines indicate event times. Gray lines are linear model fits to individual repeats.}
\label{fig:simulated_outliers}
\end{center}
\end{figure}

\citet{pinheiro_efficient_2001} discuss two 
types of longitudinal outliers: 
1) outlying individuals, 
and 2) outlying observations within individuals. Examples of outlying individuals include subjects 
with very high/low health status at baseline, 
and/or with rapid progression. In other words, outlying individuals correspond to outliers in ${\mv B}_i$. Examples of outlying observations might be a few observations that are quite 
different than the rest of the observations for a given individual. In other words, outlying observations correspond to outliers in $Z_{ij}$. Figure \ref{fig:simulated_outliers} 
displays simulated realisations from a robust joint model. As expected, most of the subjects are homogeneous, 
while a few subjects seem to have extreme trajectory or extreme observations. For instance, subjects 10, 26 and 27 seem to have relatively higher slopes compared to the rest.  
Fourth observation for subject 24 seems to 
deviate more around the individual line compared to 
the other observations for the same subject.  
Similar features can also be seen for the DIVAT data-set: 
Figure \ref{fig:spagetti_train_subset} 
illustrates subjects with high level of 
SCr at baseline, with rapid progression, 
and with a few observations that seem different 
from the rest of the observations for the same 
subjects.  

To accommodate the aforementioned outliers, the Gaussian assumption for ${\mv B}_i$ and $Z_{ij}$ 
can be relaxed using $t$ distribution that  
would give lower weights to outliers.
$t$ distribution can be specified using the variance mixtures as
${\mv B}_i = \sqrt{V_i} {\mv B}_i^{*}$ and 
$Z_{ij} = \sqrt{W_{ij}} \sigma Z_{ij}^{*}$, 
where $V_i$ and $W_{ij}$ are inverse-Gamma 
random variables, such that 
$V_i|\phi \sim \mathcal{IG}(\phi/2, \phi/2)$ and 
$W_{ij}|\delta \sim \mathcal{IG}(\delta/2, \delta/2)$, 
with the following properties, 
$V_i \perp W_{ij}$, $W_{ij} \perp W_{ij'}$ for $j \neq j'$, 
${\mv B_i^*} | {\mv \Sigma} \sim \mathcal{MVN} \left({\mv 0}, {\mv \Sigma} \right)$, and $Z_{ij}^{*} \sim \mathcal{N}(0, 1)$. 
With these specifications, one would obtain 
${\mv B_i} | {\mv \Sigma}, \phi \sim \mathcal{MV}t \left({\mv 0}, {\mv \Sigma}, \phi \right)$ 
and $Z_{ij}|\sigma, \delta \sim t(0, \sigma^2, \delta)$, 
with the following properties, 
${\mv B}_i \perp Z_{ij}$ and $Z_{ij} \perp Z_{ij'}$, 
for $j \neq j'$.  
Note that the conditionals on the mixing variates are still Normal 
such that ${\mv B}_i|V_i, {\mv \Sigma} \sim \mathcal{MVN}({\mv 0}, V_i {\mv \Sigma})$ and $Z_{ij}|\sigma, W_{ij} \sim \mathcal{N}(0, W_{ij} \sigma^2)$. 
$\phi$ and $\delta$ are the so-called degree-of-freedom 
parameters, and it is well known that if such a parameter 
converges to infinity, $t$ distribution converges to the Gaussian. 

To the best of our knowledge, there is no work in the literature that considered robust joint modelling with the above properties. \citet{taylor2013}, \citet{li_robust_2009} and \citet{huang2010}  considered robust joint 
modelling with 
Normally distributed ${\mv B}_i$ and $t$-distributed 
$Z_{ij}$ with fixed degree-of-freedom parameters. \citet{kim_sungduk2016} considered Normally distributed 
${\mv B}_i$, and generalised $t$-distributed $Z_{ij}$.  
\citet{baghfalaki2013,baghfalaki2014} considered our model with 
$W_{ij} = W_i$. Note that under their specification, 
the property, $Z_{ij} \perp Z_{ik}$ for $j \neq k$, 
does not hold, as $W_i$ is shared across $Z_{ij}$'s.  
Among these works, 
only \citet{taylor2013} considered dynamic predictions, 
whereas the rest focused on parameter estimation. 
 
\subsection{Bayesian inference}
\label{sec:inference}

In this section, we present inference for the joint model with $t$-distributed ${\mv B}_i$ and $Z_{ij}$ terms. Inference for the model with at least one of these terms being Gaussian, or $Z_{ij}$ being $t$-distributed based on $W_{ij} = W_{i}$, are just special cases. 

Let 
${\mv Y} = [Y_{1}^{\top} \ldots Y_{n}^{\top}]^{\top}$ with  
${\mv Y}_i = [Y_{i1} \ldots Y_{im_i}]^{\top}$; 
${\mv T} = [T_1 \ldots T_n ]^{\top}$;
${\mv E} = [E_1 \ldots E_n ]^{\top}$;
${\mv x} = [{\mv x}_1^{\top} \ldots {\mv x}_n^{\top}]^{\top}$ with 
${\mv x}_i = [{\mv x}_{i1} \ldots {\mv x}_{im_i}]^{\top}$ 
and ${\mv x}_{ij}$ as before; 
${\mv d} = [{\mv d}_1^{\top} \ldots {\mv d}_n^{\top}]^{\top}$ with 
${\mv d}_i = [{\mv d}_{i1} \ldots {\mv d}_{im_i}]^{\top}$ 
and ${\mv d}_{ij}$ as before; 
${\mv B} = [{\mv B}_1^{\top} \ldots {\mv B}_n^{\top}]^{\top}$ with 
${\mv B}_i$ as before; 
${\mv V} = [V_1 \ldots V_n ]^{\top}$;
${\mv W} = [{\mv W}_1^{\top} \ldots {\mv W}_n^{\top} ]^{\top}$ 
with ${\mv W}_i = [W_{i1} \ldots W_{im_i}]^{\top}$;
${\mv \zeta}$ the parameters of $h_0(t)$;
${\mv c} = [{\mv c}_1^{\top} \ldots {\mv c}_n^{\top}]^{\top}$ 
with ${\mv c}_i$ as before. 
The joint posterior density of the parameters 
and latent variables can be written as 
\begin{align}
\label{eq:joint_posterior}
f({\mv \alpha}, {\mv \Sigma}, \phi, \sigma, \delta, {\mv \zeta}, {\mv \omega},{\mv \eta}, {\mv B}, {\mv V}, {\mv W} | {\mv Y}, {\mv T}, {\mv E}, {\mv x}, {\mv d}, {\mv c}) \propto& \ f({\mv Y}|{\mv\alpha}, \sigma, {\mv B}, {\mv W}, {\mv x}, {\mv d}) \times \nonumber \\ & \ f({\mv T}, {\mv E}| {\mv \zeta}, {\mv \omega}, {\mv \eta},{\mv\alpha}, {\mv B}, {\mv x}, {\mv d}, {\mv c}) \times \nonumber \\  & 
\ f({\mv B}|{\mv V}, {\mv\Sigma}) f({\mv V}|\phi) f({\mv W}|\delta) \times  \nonumber \\  & 
\ f({\mv \alpha}, {\mv \Sigma}, \phi, \sigma, \delta, {\mv \zeta}, {\mv \omega}, {\mv \eta}),
\end{align}
with $f(\cdot)$ being a general notation for probability density 
function. The first distribution on the right-hand side of \eqref{eq:joint_posterior}, $f({\mv Y}|{\mv \alpha}, \sigma, {\mv B}, {\mv W}, {\mv x}, {\mv d})$, 
is based on the longitudinal sub-model \eqref{eq:eq_long}, 
and is constructed based on univariate Normal distributions 
such that 
\begin{equation*}
Y_{ij}|{\mv \alpha}, \sigma, {\mv B}_i, {W}_{ij}, {\mv x}_{ij}, {\mv d}_{ij} \sim \mathcal{N}({\mv x}_{ij}^{\top} {\mv \alpha} + {\mv d}_{ij}^{\top} {\mv B}_i, W_{ij} \sigma^2).
\end{equation*} 
The second term is based on the survival sub-model \eqref{eq:eq_surv}, 
and constructed by 
\begin{equation*}
h(T_i)^{E_i} S(T_i),
\end{equation*}
where $S(\cdot)$ being the survival function, 
defined as $S(t) = \mathbb{P}(T > t) = \exp\left(-\int_{0}^{T_i} h_i(v) dv \right)$. As mentioned in Section \ref{sec:distributional_assumptions}, 
$f({\mv B}|{\mv V}, {\mv \Sigma})$ is 
constructed based on $\mathcal{MVN}({\mv 0, V_i {\mv \Sigma}})$, and 
$f({\mv V}|\phi)$ and $f({\mv W}|\delta)$ are based on 
inverse-Gamma distributions,  
$\mathcal{IG}(\phi/2, \phi/2)$ and
$\mathcal{IG}(\delta/2, \delta/2)$, 
respectively.  
$f({\mv \alpha}, {\mv \Sigma}, \phi, \sigma, \delta, {\mv \zeta}, {\mv \omega}, {\mv \eta})$ corresponds to the joint prior 
distribution of the parameters. We assume independent priors for 
the parameters such that
\begin{equation}
f({\mv\alpha}, {\mv \Sigma}, \phi, \sigma, \delta, {\mv \zeta}, {\mv \omega}, {\mv \eta}) = f({\mv \alpha}) f({\mv \Sigma}) f(\phi) f(\sigma) f(\delta) f({\mv \zeta}) f({\mv \omega})f({\mv \eta}).
\end{equation}
$\alpha_h$ $(h = 2, \ldots, p)$ are given zero-mean 
Cauchy prior with 
scale parameter of 5, $\mathcal{C}(0, 5)$, 
whereas $\alpha_1$ is given $\mathcal{C}(0, 20)$.
${\mv \Sigma}$ is decomposed as ${\mv R} {\mv \Omega} {\mv R}$, 
where ${\mv R}$ is diagonal matrix of scale parameters 
that are specific to $B_{h}$ $(h = 1, \ldots, q)$, 
and ${\mv \Omega}$ is in the form of a correlation matrix. 
Elements of ${\mv R}$ are given half-Cauchy priors 
with scale of 5, $\mathcal{C}_{+}(0, 5)$, 
whereas elements of ${\mv \Omega}$ are given LKJ prior 
with the parameter of 2, $\mathcal{LKJ}(2)$. 
$\phi$ and $\delta$ are given uniform priors, between 2 and 100.  
$\sigma$ is given $\mathcal{C}_{+}(0, 5)$. 
Log-transformed elements of ${\mv \zeta}$ and 
elements of ${\mv \omega}$ and ${\mv \eta}$ 
are given 
$\mathcal{C}(0, 5)$. 

Samples from the joint posterior \eqref{eq:joint_posterior} 
are drawn using HMC \citep{neal2011}, specifically using the NUTS algorithm 
\citep{hoffman2014}, that is an adaptive version of HMC.  
Methods are implemented in the {\tt R} package {\tt robjm} (\url{github.com/ozgurasarstat/robjm}) \citep{r2018}, that internally uses the so-called HMC engine Stan \citep{carpenter2017} through the {\tt RStan} package \citep{rstan2018}.

 \section{Dynamic predictions}
 \label{sec:dynamic_prediction}
  
 \subsection{Definition}
 \label{sec:notation}

Our target of inference for a newcomer subject 
$k$     
is the prediction of 
subject-specific conditional failure  
probability between time points 
$s$ and $s + u$ given that the patient 
did not experience the event until time point $s$, 
i.e. $T_k^* > s$,  
and subject-specific data recorded 
up to and including time $s$ 
$\left({\mv Y}_k = [Y_{k1} \ldots Y_{km_k}]^{\top}, 
{\mv x}_k = [x_{k1} \ldots x_{km_k}]^{\top}, 
{\mv d}_k = [d_{k1} \ldots d_{km_k}]^{\top},
{\mv c}_k = [c_{i1} \ldots c_{ir}]^{\top} \right)$:	
\begin{align}
\label{eq:failure_prob}
\pi_{k}(s, u) &= \mathbb{P}(s < T_{k}^* \leq s + u | T_k^* > s, {\mv Y}_k, {\mv x}_k, {\mv d}_k, {\mv c}_k), &\nonumber\\&= 1 - \mathbb{P}\left(T_k^* > s + u  | T_k^* > s, {\mv Y}_k, {\mv x}_k, {\mv d}_k, {\mv c}_k \right),
\end{align} 
where $u > 0$ is called the lead-time or forecast horizon. 
The conditional survival probability in \eqref{eq:failure_prob} could be obtained as
\begin{align}
\mathbb{P}\left(T_k^* > s + u  | T_k^* > s, {\mv Y}_k, {\mv x}_k, {\mv d}_k, {\mv c}_k \right) &= \nonumber \\ 
\int_{{\mv \theta}} \int_{{\mv W}_k}  & \int_{V_k} \int_{{\mv B}_k} \frac{\mathbb{P}\left(T_k^* > s + u  | {\mv Y}_k, {\mv x}_k, {\mv d}_k, {\mv c}_k, {\mv B}_k, V_k, {\mv W}_k, {\mv \theta} \right)}{\mathbb{P}\left(T_k^* > s  | {\mv Y}_k, {\mv x}_k, {\mv d}_k, {\mv c}_k, {\mv B}_k, V_k, {\mv W}_k, {\mv \theta} \right)} \times \nonumber \\ & 
\ \ \ \ \ \ \ \ \ \ f({\mv B}_k, V_k, {\mv W}_k | T_k^* > s, {\mv Y}_k, {\mv x}_k, {\mv d}_k, {\mv c}_k, {\mv \theta}) \times \nonumber \\& 
\ \ \ \ \ \ \ \ \ \ f( {\mv \theta} | {\mv Y}, {\mv x}, {\mv d}, {\mv c}, {\mv T}, {\mv E}) d{\mv B}_k d V_k d{\mv W}_k  d{\mv \theta},
\end{align}
where ${\mv \theta}$ consists of all the parameters. 
For the a new subject, 
we would not have samples from 
$f({\mv B}_k, V_k, {\mv W}_k | T_k^* > s, {\mv Y}_k, {\mv x}_k, {\mv d}_k, {\mv c}_k, {\mv \theta})$, 
that are readily obtained during the inference step. 
We can draw the latent variables, 
$V_k, {\mv W}_k, {\mv B}_k$, from 
\begin{align}
f(V_k, {\mv W}_k, {\mv B}_k | T_k^* > s, {\mv Y}_k, {\mv x}_k, {\mv d}_k, {\mv c}_k, {\mv \theta}) \propto &  
\ f({\mv Y}_k | {\mv B}_k, {\mv W}_k, {\mv x}_k, {\mv d}_k, {\mv \theta}) \times \nonumber\\
&\ f(T_k, E_k | {\mv B}_k, {\mv x}_k, {\mv d}_k, {\mv c}_k, {\mv \theta}) \times \nonumber\\ & \ f({\mv B}_k|V_k, {\mv \theta}) f(V_k|{\mv \theta}) f({\mv W}_k|{\mv \theta}),
\end{align}  
using the MC  
samples of the parameters obtained 
from $[ {\mv \theta} | {\mv Y}, {\mv x}, {\mv d}, {\mv c}, {\mv T}, {\mv E}]$; see \eqref{eq:joint_posterior}. 
One would obtain dynamic predictions by updating  
$\pi_{k}(s, u)$ to $\pi_{k}(s^{\prime}, u)$, 
if subject $k$ is still at risk at time $s^{\prime}$ $(s^{\prime} > s)$ (such that $T_k^* > s^{\prime}$) by incorporating 
any new data recorded for her/him in the time interval 
$(s, s^{\prime}]$.

\subsection{Accuracy measures for dynamic predictions}

A prediction score, whether dynamic or not, requires good properties of discrimination and calibration in order for having a practical use in personalized medicine \citep{steyerberg_assessing_2010}. Methods to assess these properties have already been extensively published  \citep{graf_assessment_1999,gerds_consistent_2006,heagerty_time-dependent_2000} and have recently been extended to dynamic predictions \citep{schoop_quantifying_2008, blanche_quantifying_2015, Fournier2018}. 

A widely used discrimination measure is the 
Area Under the Receiver Operating 
Characteristics Curve (AUC) that 
aims to assess how the predictions 
distinguish between a patient who 
has the event from another patient who does not. 
In a dynamic prediction context, 
AUC is calculated for each landmark time-point $s$,
with a forecast horizon of $u$, 
such that 
\begin{align*}
\mbox{AUC}_{\pi}(s, u) = \mathbb{P} \left( \pi_k(s,u) > \pi_{k^{\prime}}(s,u) \big | s < T_k^* \leq s + u, T_{k^{\prime}}^* > s + u \right),
\end{align*}
where $k$ and ${k'}$ are indices 
for two randomly selected subjects. 
We use the estimator of \citet{blanche_quantifying_2015} 
for taking into account right-censoring. 
Higher AUC values indicate better discrimination. 

Brier score is one 
of the widely used measure to globally assess the prognostic performances. 
A disadvantage of the Brier score is 
that it depends on the 
marginal failure probabilities that could 
potentially take different values at 
different landmark times, 
and thus can be misleading in the dynamic prediction context.  
\citet{Fournier2018} proposed 
an $\mbox{R}^2$-type criterion 
that builds on the Brier score by 
adjusting it with the marginal failure 
probability such that  
\begin{align*}
\mbox{R}^2_{\pi}(s, u) = 1- \frac{\mbox{BS}_\pi(s, u)}{ \mbox{BS}_0(s, u)},
\end{align*}
where $\mbox{BS}_\pi(s, u)$ is 
the Brier score and 
$\mbox{BS}_0(s, u)$ the Brier score 
for the reference model that does not 
use any subject-specific information.  
Higher $\mbox{R}^2$ indicates better performance.  

As shown in \citet{Fournier2018}, 
both Brier score and the $\mbox{R}^2$-type 
criterion measure calibration and discrimination 
simultaneously. We therefore also use 
calibration plots to solely check the 
calibration properties of the predictions. The calibration is described as comparing predicted values within subgroups, 
e.g. defined from deciles of predictions for the 
DIVAT application, 
to observed event survival that is 
computed using the Kaplan-Meier method. 
From a Bayesian perspective, we have 
calculated deciles for each of element 
of the MC samples of dynamic predictions.

 \section{Application to the kidney transplantation data}
 \label{sec:application}

 \subsection{Joint modelling for the learning sample}
Following \citet{fournier_ndt}, 
the following 
joint model is fitted to the learning data: 
\begin{numcases}{}
\label{eq:eq_long_divat}
Y_{ij} = Y_i^*(t_{ij}) + Z_{ij} = \alpha_1 + t_{ij} \alpha_2 + B_{1i} + t_{ij} B_{2i} + Z_{ij},\\
\label{eq:eq_surv_divat}
\begin{array}{l} 
\displaystyle
\smash[b]h_i(t) = \lambda \nu t^{\nu - 1} \exp\Big(SCr3_i \omega_1 + AR_i \omega_2 + Age_i \omega_3 + CV_i \omega_4 +  ACI_i \omega_5 + \\
\qquad \qquad \qquad \qquad \ \  P2008_i \omega_6 + GR_i \omega_7 + \eta_1 Y_{i}^*(t) + \eta_2 \frac{\partial Y_{i}^*(t)}{\partial t} \Big)  
\end{array}&
\end{numcases} 
where $Y = \log(SCr)$, and 
$P2008$ stands for transplantation period: before 31 December 2007 
versus after January 2008. 
No baseline covariates were included in 
\eqref{eq:eq_long_divat} since SCr evolution is on the causal pathway between baseline factors and graft failure risk as discussed in \citet{fournier_ndt}. Besides, it was confirmed 
in the web supplementary material of \citet{fournier_ndt} that inclusion of baseline covariates in the longitudinal sub-model did not improve predictive performances.   
Note that the current value and rate of change 
parametrization is suggested by Figure 
\ref{fig:spagetti_train_subset}:  
increase in 
the mean SCr level is 
higher for the graft failure group 
compared to the censored group, 
and for the former group  
there is an acceleration in the SCr increase 
towards the event. 

In the following, 
we will distinguish between 
four joint models for the DIVAT data-set 
based on the following distributional assumptions: 
Normally distributed ${\mv B}_i$ and 
Normally distributed $Z_{ij}$ ($N-N$ model), 
$t$-distributed ${\mv B}_i$ and 
$t$-distributed $Z_{ij}$ ($t-t$),
Normally distributed ${\mv B}_i$ and 
$t$-distributed $Z_{ij}$ ($N-t$), 
$t$-distributed ${\mv B}_i$ and 
Normally distributed $Z_{ij}$ ($t-N$). 
Our general strategy would be to fit 
the $t-t$ model when there is evidence against Normal, 
and check the posterior summaries of the degrees-of-freedom parameters. As mentioned before, 
this approach is due to the fact that 
we do not know the source of heavy-tailedness 
in the standardised marginal residuals 
(see Figure \ref{fig:train_qq}). 
If results for any of the degrees-of-freedom 
parameters indicates that Normal 
assumption is reasonable, 
we then switch to $N-t$ or $t-N$ model.  
As will be discussed in the next paragraph, 
for the DIVAT application, $t-t$ model does not indicate 
Normality to either of the ${\mv B}_i$ or $Z_{ij}$ terms. 
Therefore, we include the $N-t$ and $t-N$ models for the 
sake of seeing the likely effects of wrongly assuming 
any of ${\mv B}_i$ or $Z_{ij}$ as Normal. 
For each of the models, 4 chains with lengths of 2,000 
were started from random initials. 
For each chain, first halves were considered as warm-up 
that results MC samples of 4,000 for each model.  
Convergence of the chains were checked 
using trace-plots, 
density plots for the chains, 
and R-hat statistic of \citet{brooks97}. 

\begin{figure}[t]
\begin{center}
\includegraphics[scale = 0.65]{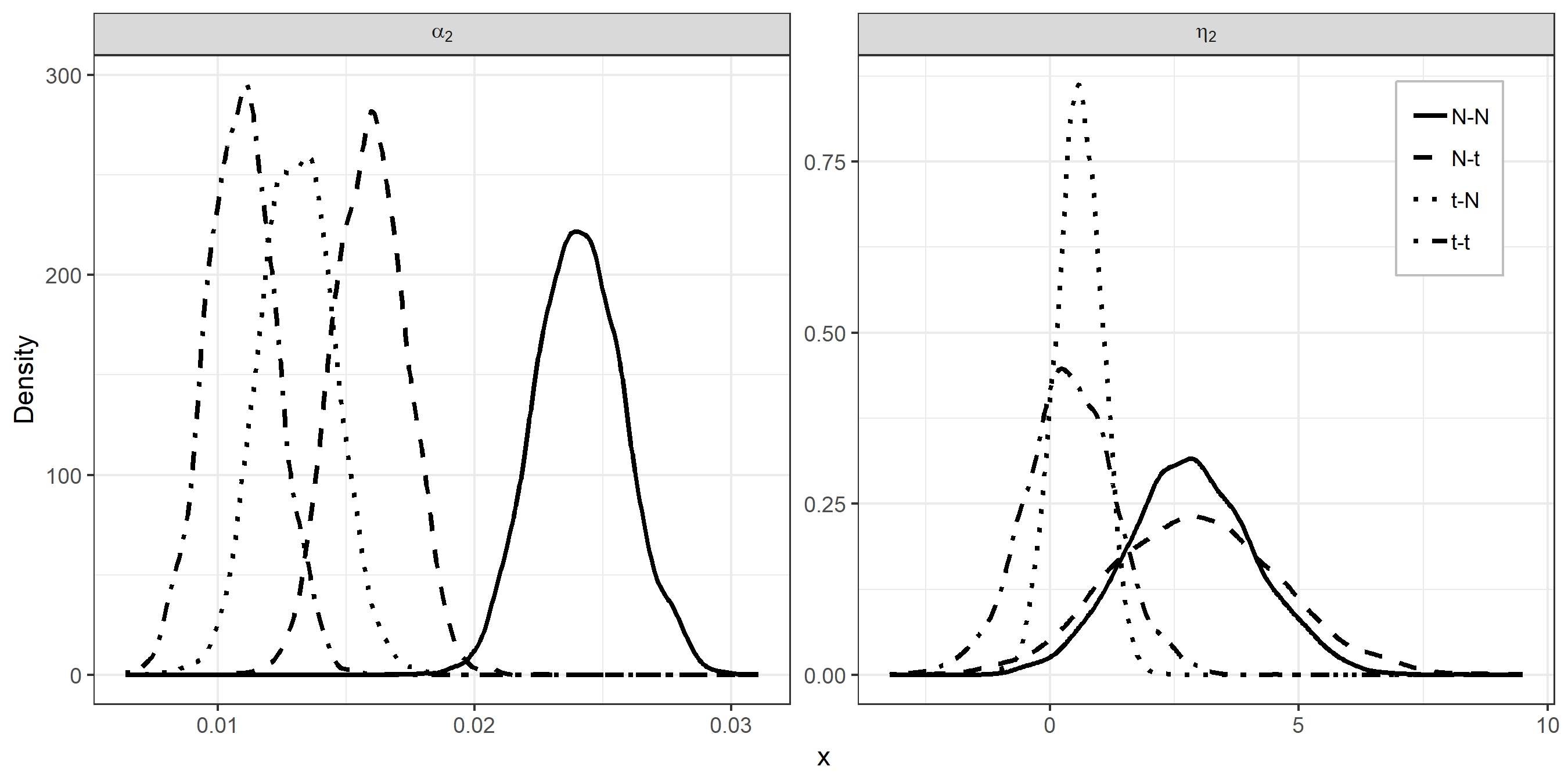}
\caption{Smoothed density plots of 
posterior samples for $\alpha_2$ (left) 
and $\eta_2$ (right) 
based on the $N-N$ (solid), $N-t$ (dashed), 
$t-N$ (dotted) and $t-t$ (dotted-dashed)  
joint models.}
\label{fig:alpha2_eta2_plot}
\end{center}
\end{figure}

Table \ref{tab:estimates} presents posterior 
summaries for each of the 4 joint models, specifically the 
2.5\%th, 50\%th and 97.5\%th percentiles, of the MC samples. 
Results for 
$\phi$ and $\delta$ under the $t-t$ model 
imply that both ${\mv B}_i$ and $Z_{ij}$ reflect 
heavier tails than the Gaussian. 
When we let both ${\mv B}_i$ and $Z_{ij}$ being Normally 
distributed (i.e. the $N-N$ model), 
the population-averaged slope, $\alpha_2$, 
is largely over-estimated, compared to the $t-t$ model. 
When we only let the error terms being 
$t$-distributed (i.e. the $N-t$ model), this 
over-estimation gets milder. 
The results of $\alpha_2$ under the $t-N$ model 
are the most 
similar to those of the $t-t$ model.  
These can be explained as the following: profiles for 
subjects who had extreme progression (high slopes) 
are better captured by the $t$-distributed ${\mv B}_i$ terms. 
However, when the error term is assumed to be 
Normal ($t-N$ model),  
all the outlying behaviours are forced to be in ${\mv B}_i$, 
hence we obtained somewhat over-estimated $\alpha_2$.   
Additionally, $\eta_2$ is largely over-estimated by the models with Normally distributed ${\mv B}_i$ regardless of  
$Z$ being Normal or $t$. 
This can also be explained by better capturing subject-specific 
slopes that in turn would results better 
predictions for $\frac{\partial Y^*_i(t)}{\partial t}$. 
Smoothed posterior densities for  
$\alpha_2$ and $\eta_2$ are presented in 
Figure \ref{fig:alpha2_eta2_plot}. 
Scatter-plots of the posterior quantiles of 
$B_{1i}$ and $B_{2i}$  
to compare the models are presented 
in Figures $1-6$ of the supplementary material. 
Note that the differences between the models are 
more apparent for $B_{2i}$. 


 \begin{table}[t]
 \begin{center}
 \caption{Posterior quantiles for the parameters based on the learning sample. 
 Reported values are the 2.5\%, 50\% and 97.5\% percentiles of the MC samples.}\vspace{0cm}
 \label{tab:estimates}
 \fbox{
 \scalebox{0.67}{
\begin{tabular}{@{\extracolsep{1pt}}l r r r r r r r r r r r r@{}} 
               & \multicolumn{3}{c}{$t-t$}  & \multicolumn{3}{c}{$N-N$}  & \multicolumn{3}{c}{$N-t$} & \multicolumn{3}{c}{$t-N$}\\ 
\cline{2-4} \cline{5-7} \cline{8-10} \cline{11-13}
               & 2.5\%  & 50\%   & 97.5\% & 2.5\%  & 50\%   & 97.5\% & 2.5\%   & 50\%    & 97.5\%   & 2.5\%   & 50\%    & 97.5\% \\ \hline
$\alpha_1$     & 4.8315 & 4.8438 & 4.8560 & 4.8463 & 4.8594 & 4.8733 & 4.8473  & 4.8612  & 4.8732   & 4.8244   & 4.8367 & 4.8488 \\
$\alpha_2$     & 0.0082 & 0.0109 & 0.0136 & 0.0208 & 0.0241 & 0.0277 & 0.0132  & 0.0159  & 0.0187   & 0.0103   & 0.0131 & 0.0159 \\
$\Sigma_{11}$  & 0.0567 & 0.0633 & 0.0699 & 0.1007 & 0.1067 & 0.1135 & 0.0973  & 0.1037  & 0.1105   & 0.0483   & 0.0535 & 0.0589 \\
$\Sigma_{12}$  &-0.0003 & 0.0004 & 0.0010 & 0.0003 & 0.0015 & 0.0028 & -0.0001 & 0.0008  & 0.0019   & 0.0000003& 0.0006 & 0.0013 \\
$\Sigma_{22}$  & 0.0013 & 0.0015 & 0.0017 & 0.0032 & 0.0036 & 0.0040 & 0.0020  & 0.0022  & 0.0025   & 0.0013   & 0.0015 & 0.0018 \\
$\sigma^2$     & 0.0067 & 0.0072 & 0.0076 & 0.0108 & 0.0194 & 0.0200 & 0.0066  & 0.0071  & 0.0075   & 0.0176   & 0.0182 & 0.0188 \\
$\phi$         & 3.2205 & 3.8003 & 4.5526 &        &        &        &         &         &          & 2.4977   & 2.7930 & 3.1372 \\
$\delta$       & 2.7149 & 2.9099 & 3.1365 &        &        &        & 2.5562  & 2.7284  & 2.9269   &          &        & \\
$\log(\lambda)$&-22.1580&-20.3039&-18.5817&-22.7288&-20.7840&-18.9112&-21.3376 &-19.4082 & -17.4713 & -22.6612 &-20.9944& -19.3376\\
$\log(\nu)$    & 0.2683 & 0.3543 & 0.4364 & 0.2323 & 0.3262 & 0.4145 & 0.2536  & 0.3407  & 0.4244   & 0.2643   & 0.3536 & 0.4407 \\
$\omega_1$     &-0.3190 &-0.2060 &-0.0959 &-0.3029 &-0.1897 &-0.0786 & -0.2764 & -0.1661 & -0.0597  & -0.3052  & -0.2022& -0.1005\\
$\omega_2$     & 0.1471 & 0.3717 & 0.5941 & 0.1400 & 0.3735 & 0.5903 & 0.1669  & 0.3854  & 0.6048   & 0.1342   & 0.3668 & 0.5908 \\
$\omega_3$     & 0.2662 & 0.3719 & 0.4902 & 0.2857 & 0.3955 & 0.5068 & 0.2503  & 0.3578  & 0.4679   & 0.2976   & 0.4098 & 0.5184 \\
$\omega_4$     & 0.1497 & 0.3425 & 0.5337 & 0.1474 & 0.3409 & 0.5357 & 0.1394  & 0.3372  & 0.5345   & 0.1626   & 0.3596 & 0.5520 \\
$\omega_5$     & 0.2237 & 0.4575 & 0.6849 & 0.1977 & 0.4354 & 0.6589 & 0.2200  & 0.4433  & 0.6618   & 0.2289   & 0.4677 & 0.6949 \\
$\omega_6$     &-0.5028 &-0.2591 &-0.0112 &-0.5435 &-0.2987 &-0.0612 & -0.5218 & -0.2814 &-0.0307   & -0.5598  & -0.3085& -0.0558 \\
$\omega_7$     &-0.0114 & 0.2515 & 0.5169 &-0.0042 & 0.2688 & 0.5334 & -0.0231 & 0.2507  & 0.5154   & -0.0183  & 0.2464 & 0.5100 \\
$\eta_1$       & 2.6302 & 2.9904 & 3.3574 & 2.6364 & 3.0395 & 3.4403 & 2.3990  & 2.8054  & 3.2102   & 2.7554   & 3.0792 & 3.4055 \\
$\eta_2$       &-1.4585 & 0.3872 & 2.2447 & 0.3784 & 2.8103 & 5.3856 & -0.2481 & 2.8887  & 6.4138   & -0.3726  & 0.5769 & 1.4432 \\
 \end{tabular}}
 }
 \end{center}
 \end{table}

 \subsection{Dynamic predictions for the validation sample}

 \subsubsection{Accuracy measures}

 Dynamic predictions are calculated at six 
 post-transplantation landmark times:  
 $s = 1, 2, 3, 4, 5, 6$. 
 A clinically meaningful forecast window of 
 5 years is considered, i.e. $u = 5$.  
 Table \ref{tab:landmark_subjects} is 
 the frequency table regarding 
 subjects who were at risk at the landmark times, 
 subjects who had the event, and 
 who were censored within the forecast horizons, 
 and subjects who survived beyond the forecast horizons. 
 \begin{table}[t]
 \begin{center}
 \caption{Number of patients 
 (and associated percentages) 
 who were at risk at the landmark times (nrisk),  
 who had graft failure (nevent) and 
 were censored (ncens) in the 5-year forecast horizon, 
 and survived beyond the 5-year forecast horizon (nsurv). 
 } \vspace{0.3cm}
 \label{tab:landmark_subjects}
 \fbox{
 \scalebox{1}{
\begin{tabular}{@{\extracolsep{1pt}}r r r r r@{}} 
$s$ & nrisk & nevent       & ncens          & nsurv\\ \hline
1   & 2,523 & 291 (11.5\%) & 1,451 (57.5\%) & 781 (31.0\%) \\
2   & 1,796 & 262 (14.6\%) & 908   (50.6\%) & 626 (34.9\%)\\
3   & 1,459 & 268 (15.6\%) & 722   (49.5\%) & 509 (34.9\%)\\
4   & 1,157 & 188 (16.2\%) & 585   (50.6\%) & 384 (33.2\%) \\
5   & 950   & 161 (16.9\%) & 529   (55.7\%) & 260 (27.4\%)\\ 
6   & 781   & 123 (15.7\%) & 480   (61.5\%) & 178 (22.8\%) \\ 
 \end{tabular}}
 }
 \end{center}
 \end{table}
 
 Posterior summaries of the AUC and R$^2$ are 
 displayed in Figure \ref{fig:acc_plots}. 
 Calibration plots are dislayed 
 in Figures $7-10$ of the supplementary material, 
 whereas the estimated slopes fit to 
 predicted risk versus observed risk are reported in 
 Table \ref{tab:calibration_slopes}.  
 Note that as mentioned in \citet{Fournier2018}, 
 $\mbox{R}^2$ can take negative values.  
 Globally, discrimination is good for all the models,  
 AUC's increase with landmark times, and 
 calibration performances appear reasonable. 
 More interestingly, comparing the models, 
 we observe the best results under the $t-N$ model 
 in terms of AUC and $R^2$. 
 It is followed by the $t-t$ model and 
 the $N-t$ model. 
 The $N-N$ has the worst performance. 
 In terms of calibration, 
 again the $t-N$ and $t-t$ models appeared the best 
 two. 
 Whereas in landmark times $1-3$, $t-t$ model 
 had higher slope estimates compared to the $t-N$ model, 
 for the later times the estimates are almost the same. 
 
 We also calculated 2.5th, 50th and 97.5th percentiles 
 of the dynamic predictions for each individual 
 for each landmark times. 
 Scatter-plots of these statistics are displayed 
 in Figures $11-16$ of the supplementary material.   
 It can be seen, 
 e.g. based on the medians, 
 that there are some subjects 
 for whom the models quite disagree, 
 e.g. see the dots that are above the $x = y$ lines. 
 One can deduce that 
 there are some subjects (e.g. outlying subjects) 
 whose dynamic predictions benefit 
 from the robust joint model with $t$ distributions, 
 especially $t$-distributed ${\mv B}_i$ terms. 
\begin{figure}[t]
\begin{center}
\includegraphics[scale = 0.65]{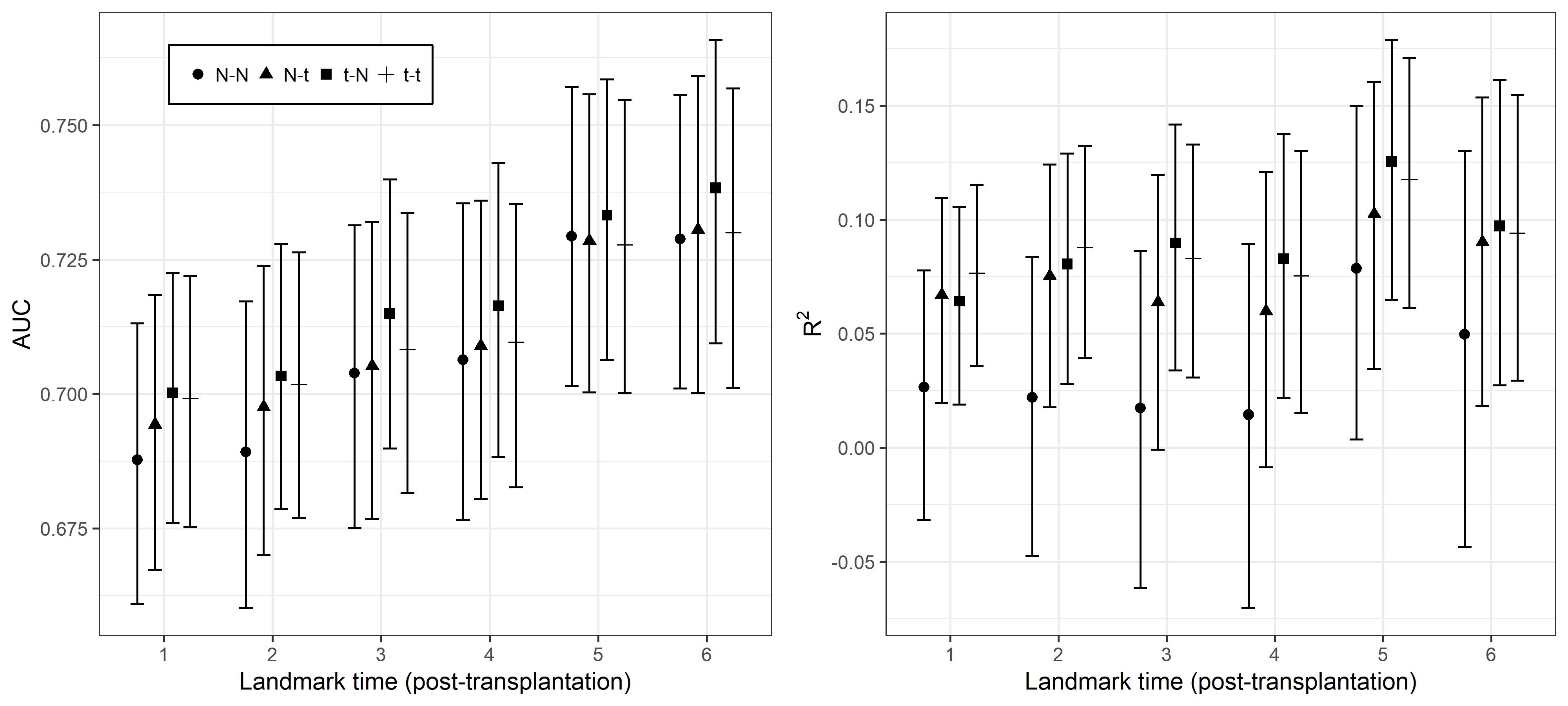}
\caption{AUC and R$^2$ across the landmark time-points. 
Middle points are the medians, whereas 
the accompanying lower and upper bars 
are the 2.5th and 97.5th percentiles 
for the accuracy measures, respectively.}
\label{fig:acc_plots}
\end{center}
\end{figure} 
   
 \begin{table}[t]
 \begin{center}
 \caption{Slope estimates based on the linear model fits to 
 predicted risk versus observed risk from the calibration 
 statistics.} \vspace{0.3cm}
 \label{tab:calibration_slopes}
 \fbox{
 \scalebox{1}{
\begin{tabular}{@{\extracolsep{1pt}}r r r r r@{}} 
$s$ & $N-N$ & $N-t$ & $t-N$ & $t-t$\\ \hline
1   & 0.48  & 0.56  & 0.58  & 0.61 \\
2   & 0.46  & 0.55  & 0.57  & 0.60 \\
3   & 0.46  & 0.53  & 0.55  & 0.57 \\
4   & 0.48  & 0.53  & 0.55  & 0.55 \\
5   & 0.56  & 0.59  & 0.64  & 0.63 \\ 
6   & 0.52  & 0.58  & 0.60  & 0.60 \\ 
 \end{tabular}}
 }
 \end{center}
 \end{table}

%
%

 \subsubsection{Results for two patients}
\begin{figure}[t]
\begin{center}
\includegraphics[scale = 0.65]{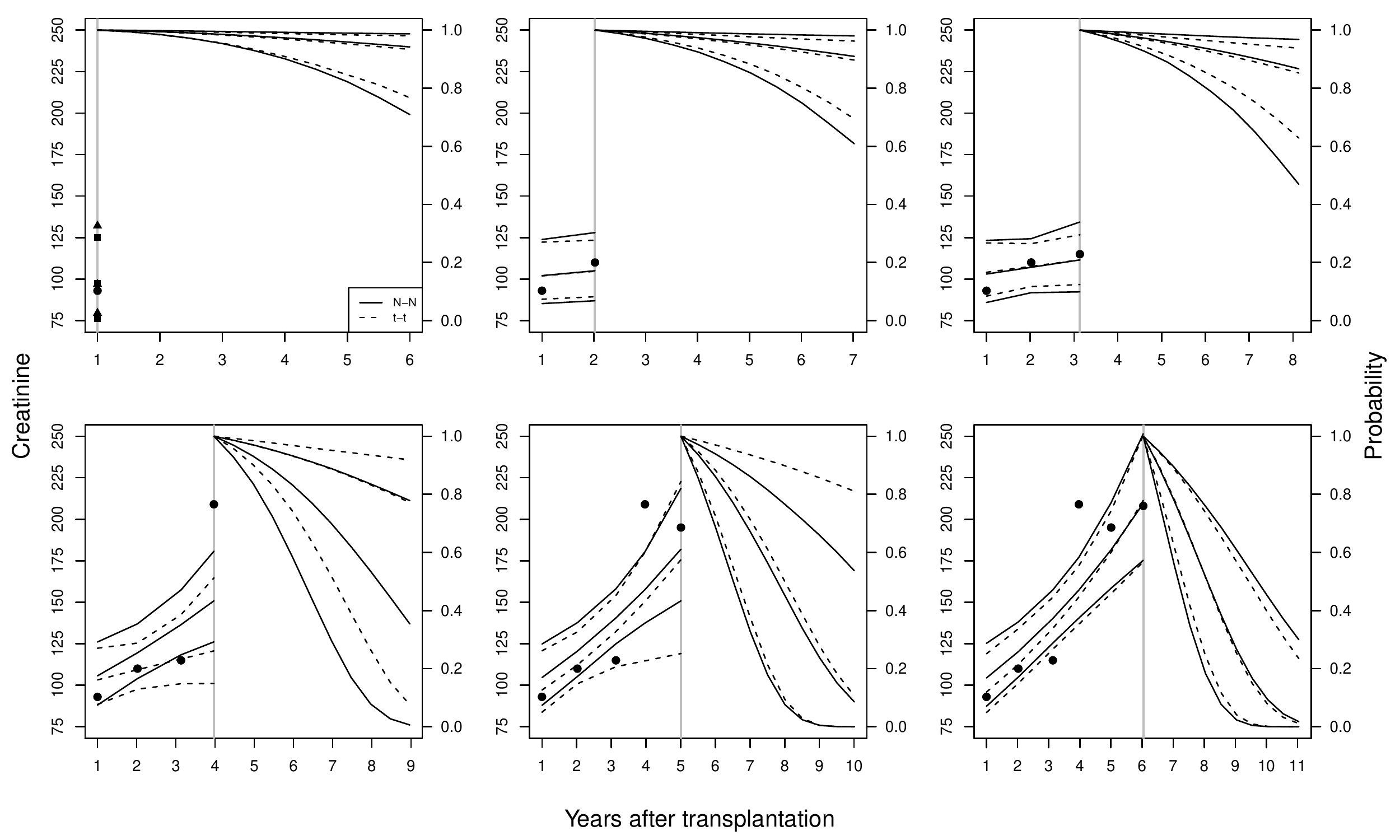}
\caption{Dynamic predictions for the first person 
from the validation sample.  
Dots are observed data, 
solid black lines are the results based on the 
standard joint model with Gaussian distributions, 
dashed based on the robust joint model with $t$ distributions.
Lower, mid and upper lines are the 2.5th, 50th and 97.5th 
percentiles of the Monte Carlo samples, respectively. 
Vertical gray line 
is the landmarking time. 
Predictions on the left-hand 
side of it are for serum creatinine, whereas on the right-hand 
side are for forecast probabilities.  
For the top-left panel, the lines on the left-hand 
side are replaced by squares (for Normal) and 
triangles (for $t$). 
See text for the patient characteristics.}
\label{fig:prob_plot_1324}
\end{center}
\end{figure}

\begin{figure}[t]
\begin{center}
\includegraphics[scale = 0.65]{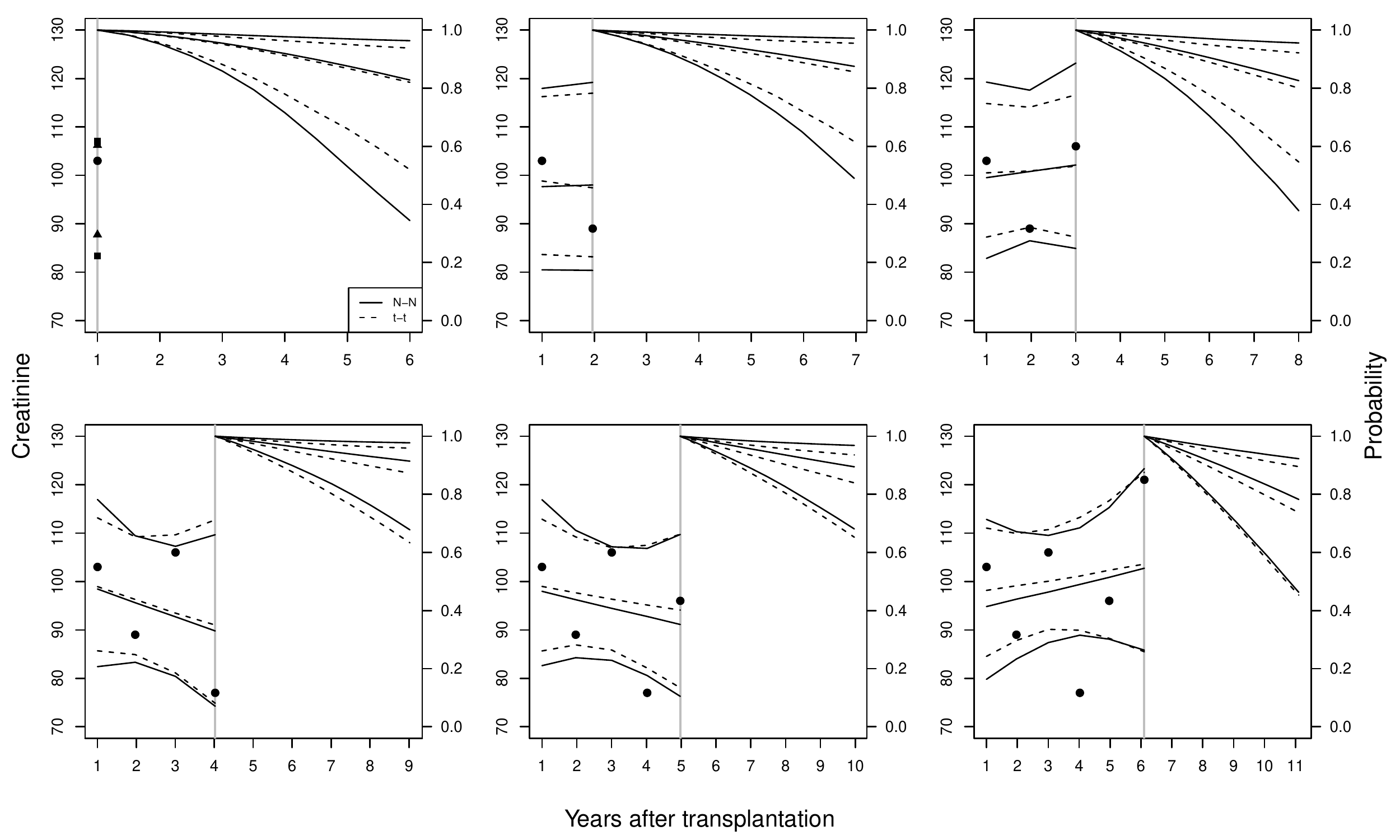}
\caption{Dynamic predictions for the 
second person.  
For explanations, see the caption of 
Figure \ref{fig:prob_plot_1324}, 
the text for the patient characteristics.}
\label{fig:prob_plot_488}
\end{center}
\end{figure}

 We present 5-year dynamic predictions for two patients 
 from the validation sample. 
 They are the same subjects presented in \citet{fournier_ndt}. 
 The first patient (Figure \ref{fig:prob_plot_1324}) was 
 a female, aged 51 years, transplanted in 2005 for the first time, without history of a cardiovascular disease, immunized against HLA class I, with a SCr measurement at 3 
 months post-transplantation of 88 $\mu$mol/L, 
 and without acute rejection episode in the first year post-transplantation. 
 The recipient was returned to dialysis at 9.34 years after
 transplantation.  
 The second patient (Figure \ref{fig:prob_plot_488}) was  
 a female, aged 60 years at transplantation, transplanted in 2007 for the second time, 
 without history of a cardiovascular 
 disease, immunized against HLA class I, 
 and a SCr measurement at 3 months post-transplantation of 
 100 $\mu$mol/L, and at least one acute rejection 
 episode in the first year after transplantation. 
 The recipient was still alive with a functioning 
 graft at 10.14 years post-transplantation.  
 Figures \ref{fig:prob_plot_1324} \& \ref{fig:prob_plot_488} 
 only present the results under the $N-N$ and $t-t$ models, 
 whereas results under all the four models 
 are presented in Figures 17 \& 18 of the supplementary 
 material. 
 For these patients, 
 dynamic predictions obtained under the 
 models generally agree. 
 There is a notable difference for the first patient 
 in the fourth landmark time-point; see 
 mid-right panel of Figure \ref{fig:prob_plot_1324}. 
 The difference is due to the fourth SCr measurement 
 that is a typical example of outlying observations:   
 the fourth $SCr$ was 209, whereas 
 the first three were 93, 110, and 115.  
 $N-N$ and $t-N$ assumptions overreact 
 and produce 
 considerably lower survival probabilities 
 compared to the $N-t$ and $t-t$. 
 Median (2.5\% and 97.5\% percentiles of the MC samples) of the 
 survival probabilities for 8.97 years after transplantation, i.e. 
 $\mathbb{P}(T_k^* > 8.97 | 
 T_k^* > 3.97, Y_k = \{\log(93), \log(110), \log(115), \log(209)\}, 
 \mv{x}_k, \mv{d}_k, \mv{c}_k)$, were 
 0.35 (0.01, 0.78), 
 0.35 (0.00002, 0.82), 
 0.74 (0.27, 0.92),
 0.77 (0.08, 0.92)
 under $N-N$, $t-N$, $N-t$ and $t-t$ 
 models, respectively. 
 Note that the patient had graft failure at 
 year 9.3, i.e. $T_k^* = 9.3$.  
 The fifth $SCr$ measurement for the 
 same subject is 195 that is closer in magnitude to the 
 fourth measurement compared to the first three. 
 $N-N$, $t-N$ and $t-t$ models correctly react to 
 the fifth measurement and update the predictions accordingly.   
 However, it $N-t$ model seems to produce somewhat higher 
 median probabilities compared to these three models. 
 
  
 \section{Discussion}
 \label{sec:discussion}

In this work, we considered dynamic predictions of kidney graft
survival using joint modelling of longitudinal and survival outcomes. 
We mainly focused on how the distributional assumptions 
would impact the predictions: widely used Normal distribution was 
compared against $t$. The proposed joint model with $t$ 
distributional assumptions is novel, 
as no work in the literature considered 
such a general model. 
Bayesian methods were 
considered for estimation and dynamic predictions. 
The proposed methods are implemented in the {\tt R} package 
{\tt robjm}. Methods are applied to data for kidney transplant patients from the French cohort DIVAT. 
Impacts of distributional assumptions 
on dynamic prediction performances
were inspected through accuracy measures 
and predictions on two individuals are illustrated. 

Regarding the DIVAT data-set, 
degree-of-freedom 
results indicated that there are both 
outlying individuals and outlying observations. 
We observed important differences between the 
standard and the robust joint models. 
The population averaged slope in the 
mixed-effects sub-model and the association parameter 
for the individual rate of change in the 
survival sub-model were largely 
over-estimated by the Gaussian joint model.  
In terms of dynamic predictions, we observed 
better calibration and discrimination 
for the robust joint models. 
Intermediate joint models, 
i.e. the $N-t$ and $t-N$ assumptions, 
suggest that outlying subjects have greater impact on dynamic predictions if they are not considered compared to outlying observations. 
Regarding individual dynamic predictions, 
robust models produced better results for the 
patient with outlying observation, 
whereas the two models  
produced similar results for the 
subject with no outlying observation.
 
Our current work voluntarily does not include a simulation study. Indeed, the objective is to present a case-study on 
predicting kidney graft failure risk,  
specifically for the patients in the DIVAT cohort. 
We clearly illustrated that considering longitudinal 
outliers have an impact on prognostic accuracy in the kidney transplantation context. Our modelling strategies may be applicable to others clinical contexts. When comparing the four joint models, we also voluntarily do not consider metrics for model selection. Indeed, since we are mainly interested in dynamic predictions, we compare 
the models using predictive accuracy measures 
obtained from a validation data-set. 

In our robust joint model proposal, we considered symmetric $t$ distribution 
as an alternative to the Gaussian. 
Given that distributional assumptions 
might have considerable impacts on dynamic predictions, 
it would be worth to study other distributions than 
symmetric $t$, e.g. skew-$t$.  
We considered joint modelling framework with the 
shared-parameter formulation to link the longitudinal and 
survival sub-models. Performances of other methods, 
e.g. latent class joint modelling \citep{proust_lima2014}, 
Cox model with time-varying covariates, 
or landmarking methods \citep{van_houwelingen_dynamic_2007}, 
in the presence of longitudinal outliers would also 
be interesting to investigate.   
Diagnostics tools for 
checking the appropriateness of 
distributional assumptions 
for non-Gaussian joint models 
would be interesting. In this study, 
this was secondary to us, 
since our aim was to inspect 
if we can improve the dynamic predictions obtained 
from the widely used Normal assumption, 
and checked this using accuracy measures 
for the validation sample. 

\citet{fournier_ndt} previously presented 
dynamic predictions for the DIVAT patients 
using the $N-N$ joint model with a frequentist 
point of view. We are able to re-produce their 
results using our $N-N$ model. 
The AUC and $R^2$ values presented in \citep{fournier_ndt} 
are higher than those presented in the current paper. 
The reason for this is that they used median of the 
MC samples to obtain the point estimates and 
frequentist methods to obtain the associated intervals 
rather than obtaining MC samples of the accuracy measures. 
Yet, following their approach, we are able to obtain 
the values they presented.

In conclusion, this study presents 
improved dynamic predictions of kidney graft failure based on  
robust joint modelling framework. 
The most important component 
to obtain improved predictions 
seems the random-effects terms. 
Nonetheless, even letting only the 
error term being $t$-distributed 
improves predictions compared to the Gaussian model. 
We prefer working with the $t-t$ model, since i) it consists of the $N-t$ and $t-N$ models as special 
cases,
ii) it is one of the best in terms of accuracy measures, 
iii) it provides a good compromise between the two special 
cases as presented for the first patient.  
The predictions based on the $t-t$ model  
will be   
deployed into the {\tt DynPG} {\tt Shiny} application 
of \cite{fournier_ndt} 
(available at {\tt https://shiny.idbc.fr/DynPG}) 
that is currently based on Gaussian joint modelling. 
This would allow physicians and patients 
get benefit from our methods. 

 \section*{Acknowledgements}
Dr. Özgür Asar was funded by the French Embassy to Turkey for a visit to The University of Nantes, which enabled great progress for the current work. Helpful discussions with 
Dr. Jonas Wallin (Lund University), 
Prof. Peter Diggle (Lancaster University) and 
Dr. Yohann Foucher (Nantes University) 
are greatfully acknowledged. 
The analysis and interpretation of data collected from the French DIVAT (Données Informatisées et VAlidées en Transplantation) cohort* (www.divat.fr, No. CNIL 914184) are the responsibility of the authors. We wish to thank members of the data manager of the DIVAT cohort (Clarisse Kerleau) and clinical research assistant team (S. Le Floch, A. Petit, J. Posson, C. Scellier, V. Eschbach, K. Zurbonsen, C. Dagot, F. M’Raiagh, V. Godel, X. Longy, P. Przednowed). We are also grateful to Roche Pharma, Novartis and Sanofi laboratories for supporting the DIVAT cohort as the CENTAURE foundation (www.fondation-centaure.org). 

* DIVAT cohort collaborators (Medical Doctors, Surgeons, HLA Biologists):
Nantes: G. Blancho, J. Branchereau, D. Cantarovich, A. Chapelet, J. Dantal, C. Deltombe, L. Figueres, C. Garandeau, M. Giral, C. Gourraud-Vercel, M. Hourmant, G. Karam, C. Kerleau, A. Meurette, S. Ville, C. Kandell, A. Moreau, K. Renaudin, A. Cesbron, F. Delbos, A. Walencik, A. Devis; Paris-Necker: L. Amrouche, D. Anglicheau, O. Aubert, L. Bererhi, C. Legendre, A. Loupy, F. Martinez, R. Sberro-Soussan, A. Scemla, C. Tinel, J. Zuber; Nancy: P. Eschwege, L. Frimat, S. Girerd, J. Hubert, M. Ladriere, E. Laurain, L. Leblanc, P. Lecoanet, J-L. Lemelle; Lyon E. Hériot: L. Badet, M. Brunet, F. Buron, R. Cahen, S. Daoud, C. Fournie, A. Grégoire, A. Koenig, C. Lévi, E. Morelon, C. Pouteil-Noble, T. Rimmelé, O. Thaunat; Montpellier: S. Delmas, V. Garrigue, M. Le Quintrec, V. Pernin, J-E. Serre.



\begin{thebibliography}{36}
\providecommand{\natexlab}[1]{#1}
\providecommand{\url}[1]{\texttt{#1}}
\expandafter\ifx\csname urlstyle\endcsname\relax
  \providecommand{\doi}[1]{doi: #1}\else
  \providecommand{\doi}{doi: \begingroup \urlstyle{rm}\Url}\fi

\bibitem[Asar et~al.(2015)Asar, Ritchie, Kalra, and Diggle]{asar2015}
Ö. Asar, J.~Ritchie, P.~A. Kalra, and P.~J. Diggle.
\newblock Joint modelling of repeated measurement and time-to-event data: an
  introductory tutorial.
\newblock \emph{International Journal of Epidemology}, 44\penalty0
  (1):\penalty0 334--344, 2015.

\bibitem[Baghfalaki et~al.(2013)Baghfalaki, Ganjali, and
  Berridge]{baghfalaki2013}
T.~Baghfalaki, M.~Ganjali, and D.~Berridge.
\newblock Robust joint modeling of longitudinal measurements and time to event
  data using normal/independent distributions: A bayesian approach.
\newblock \emph{Biometrical Journal}, 55\penalty0 (6):\penalty0 844--865, 2013.

\bibitem[Baghfalaki et~al.(2014)Baghfalaki, Ganjali, and
  Hashemi]{baghfalaki2014}
T.~Baghfalaki, M.~Ganjali, and R.~Hashemi.
\newblock Bayesian joint modeling of longitudinal measurements and
  time-to-event data using robust distributions.
\newblock \emph{Journal of Biopharmaceutical Statistics}, 24\penalty0
  (4):\penalty0 834--185, 2014.

\bibitem[Blanche et~al.(2015)Blanche, {Proust-Lima}, Loub\`ere, Berr,
  Dartigues, and {Jacqmin-Gadda}]{blanche_quantifying_2015}
P.~Blanche, C.~{Proust-Lima}, L.~Loub\`ere, C.~Berr, J-F. Dartigues, and
  H.~{Jacqmin-Gadda}.
\newblock Quantifying and comparing dynamic predictive accuracy of joint models
  for longitudinal marker and time-to-event in presence of censoring and
  competing risks.
\newblock \emph{Biometrics}, 71\penalty0 (1):\penalty0 102--113, 2015.

\bibitem[Brooks and Gelman(1997)]{brooks97}
S.~P. Brooks and A.~Gelman.
\newblock General methods for monitoring convergence of iterative simulations.
\newblock \emph{Journal of Computational and Graphical Statistics}, 7:\penalty0
  434--455, 1997.

\bibitem[Carpenter et~al.(2017)Carpenter, Gelman, Hoffman, Lee, Goodrich,
  Betancourt, Brubaker, Guo, Li, and Riddell]{carpenter2017}
B.~Carpenter, A.~Gelman, M.~D. Hoffman, D.~Lee, B.~Goodrich, M.~Betancourt,
  M.~Brubaker, J.~Guo, P.~Li, and A.~Riddell.
\newblock Stan: A probabilistic programming language.
\newblock \emph{Journal of Statistical Software}, 76\penalty0 (1):\penalty0
  1--32, 2017.

\bibitem[Cox(1972)]{cox1972}
D.~R. Cox.
\newblock Regression models and life-tables.
\newblock \emph{Journal of the Royal Statistical Society. Series B
  (Methodological)}, 34\penalty0 (2):\penalty0 187--220, 1972.

\bibitem[Flores et~al.(2013)Flores, Glusman, Brogaard, Price, and
  Hood]{Flores2013}
M.~Flores, G.~Glusman, K.~Brogaard, N.~D. Price, and L.~Hood.
\newblock P4 medicine: How systems medicine will transform the healthcare
  sector and society.
\newblock \emph{Personalized Medicine}, 10\penalty0 (6):\penalty0 565--576,
  2013.

\bibitem[Fournier et~al.(2016)Fournier, Foucher, Blanche, Buron, Giral, and
  Dantan]{Fournier2016c}
M.-C. Fournier, Y.~Foucher, P.~Blanche, F.~Buron, M.~Giral, and E.~Dantan.
\newblock A joint model for longitudinal and time-to-event data to better
  assess the specific role of donor and recipient factors on long-term kidney
  transplantation outcomes.
\newblock \emph{European Journal of Epidemiology}, 31\penalty0 (5):\penalty0
  469--479, 2016.

\bibitem[Fournier et~al.(2018)Fournier, Dantan, and Blanche]{Fournier2018}
M.-C. Fournier, E.~Dantan, and P.~Blanche.
\newblock An {{R2}}-curve for evaluating the accuracy of dynamic predictions.
\newblock \emph{Statistics in Medicine}, 37\penalty0 (7):\penalty0 1125--1133,
  2018.

\bibitem[Fournier et~al.(2019)Fournier, Foucher, Blanche, Legendre, Girerd,
  Ladri\'ere, Morelon, Buron, Rostaing, Kamar, Mourad, Garrigue, Giral, and
  Dantan]{fournier_ndt}
M.-C. Fournier, Y.~Foucher, P.~Blanche, C.~Legendre, S.~Girerd, M.~Ladri\'ere,
  E.~Morelon, F.~Buron, L.~Rostaing, N.~Kamar, G.~Mourad,
  V.~Couvrat-Desvergnes~G. Garrigue, M.~Giral, and E.~Dantan.
\newblock Dynamic predictions of long-term kidney graft failure: An information
  tool promoting patient-centered care.
\newblock \emph{Nephrology Dialysis Transplantation}, pages 1--9, 2019.
\newblock \doi{10.1093/ndt/gfz027}.

\bibitem[Gerds and Schumacher(2006)]{gerds_consistent_2006}
T.~A. Gerds and M.~Schumacher.
\newblock Consistent {{Estimation}} of the {{Expected Brier Score}} in
  {{General Survival Models}} with {{Right}}-{{Censored Event Times}}.
\newblock \emph{Biometrical Journal}, 48\penalty0 (6):\penalty0 1029--1040,
  2006.

\bibitem[Graf et~al.(1999)Graf, Schmoor, Sauerbrei, and
  Schumacher]{graf_assessment_1999}
E.~Graf, C.~Schmoor, W.~Sauerbrei, and M.~Schumacher.
\newblock Assessment and comparison of prognostic classification schemes for
  survival data.
\newblock \emph{Statistics in Medicine}, 18\penalty0 (17-18):\penalty0
  2529--2545, 1999.

\bibitem[Heagerty et~al.(2000)Heagerty, Lumley, and
  Pepe]{heagerty_time-dependent_2000}
P.~J. Heagerty, T.~Lumley, and M.~S. Pepe.
\newblock Time-dependent {{ROC}} curves for censored survival data and a
  diagnostic marker.
\newblock \emph{Biometrics}, 56\penalty0 (2):\penalty0 337--44, 2000.

\bibitem[Hoffman and Gelman(2014)]{hoffman2014}
M.~D. Hoffman and A.~Gelman.
\newblock The no-u-turn sampler: adaptively setting path lengths in hamiltonian
  monte carlo.
\newblock \emph{Journal of Machine Learning Research}, 15:\penalty0 1593--1623,
  2014.

\bibitem[Huang et~al.(2010)Huang, Li, and Elashoff]{huang2010}
X.~Huang, G.~Li, and R.~M. Elashoff.
\newblock A joint model of longitudinal and competing risks survival data with
  heterogeneous random effects and outlying longitudinal measurements.
\newblock \emph{Statistics and Its Interface}, 3:\penalty0 185--195, 2010.

\bibitem[Kaboré et~al.(2017)Kaboré, Haller, Harambat, Heinze, and
  Leffondré]{Kabore2017}
R.~Kaboré, M.~C. Haller, J.~Harambat, G.~Heinze, and K.~Leffondré.
\newblock Risk prediction models for graft failure in kidney transplantation: A
  systematic review.
\newblock \emph{Nephrology Dialysis Transplantation}, 32:\penalty0 ii68--ii76,
  2017.

\bibitem[Lange et~al.(1989)Lange, Little, and Taylor]{lange_robust_1989}
K.~L. Lange, R.~J.~A. Little, and J.~M.~G. Taylor.
\newblock Robust {{Statistical Modeling Using}} the t {{Distribution}}.
\newblock \emph{Journal of the American Statistical Association}, 84\penalty0
  (408):\penalty0 881--896, 1989.

\bibitem[Li et~al.(2009)Li, Elashoff, and Li]{li_robust_2009}
N.~Li, R.~M. Elashoff, and G.~Li.
\newblock Robust {{Joint Modeling}} of {{Longitudinal Measurements}} and
  {{Competing Risks Failure Time Data}}.
\newblock \emph{Biometrical Journal}, 51\penalty0 (1):\penalty0 19--30, 2009.

\bibitem[Neal(2011)]{neal2011}
R.~Neal.
\newblock Mcmc using hamiltonian dynamics.
\newblock In S.~Brooks, A.~Gelman, G.~L. Jones, and X.L. Meng, editors,
  \emph{Handbook of Markov Chain Monte Carlo}, pages 113--162. Chapman \&
  Hall/CRC Press, Boca Raton, 2011.

\bibitem[Pinheiro et~al.(2001)Pinheiro, Liu, and Wu]{pinheiro_efficient_2001}
J.C. Pinheiro, C.~Liu, and Y.~N. Wu.
\newblock Efficient {{Algorithms}} for {{Robust Estimation}} in {{Linear
  Mixed}}-{{Effects Models Using}} the {{Multivariate}} t {{Distribution}}.
\newblock \emph{Journal of Computational and Graphical Statistics}, 10\penalty0
  (2):\penalty0 249--276, 2001.

\bibitem[{Proust-Lima} and Blanche(2014)]{proust-lima_dynamic_2014}
C.~{Proust-Lima} and P.~Blanche.
\newblock Dynamic {{Predictions}}.
\newblock In \emph{Wiley {{StatsRef}}: {{Statistics Reference Online}}}. {John
  Wiley \& Sons, Ltd}, 2014.

\bibitem[{Proust-Lima} and Taylor(2009)]{proust-lima_development_2009}
C.~{Proust-Lima} and J.~M.~G. Taylor.
\newblock Development and validation of a dynamic prognostic tool for prostate
  cancer recurrence using repeated measures of posttreatment {{PSA}}: A joint
  modeling approach.
\newblock \emph{Biostatistics}, 10\penalty0 (3):\penalty0 535--549, 2009.

\bibitem[Proust-Lima et~al.(2014)Proust-Lima, Séne, Taylor, and
  Jacqmin-Gadda]{proust_lima2014}
C.~Proust-Lima, M.~Séne, J.~M. Taylor, and H.~Jacqmin-Gadda.
\newblock Joint latent class models for longitudinal and time-to-event data: A
  review.
\newblock \emph{Statistical Methods in Medical Research}, 23\penalty0
  (1):\penalty0 70--90, 2014.

\bibitem[{R Core Team}(2018)]{r2018}
{R Core Team}.
\newblock \emph{R: A Language and Environment for Statistical Computing}.
\newblock R Foundation for Statistical Computing, Vienna, Austria, 2018.
\newblock URL \url{https://www.R-project.org/}.

\bibitem[Rizopoulos(2011)]{rizopoulos_dynamic_2011}
D.~Rizopoulos.
\newblock Dynamic predictions and prospective accuracy in joint models for
  longitudinal and time-to-event data.
\newblock \emph{Biometrics}, 67\penalty0 (3):\penalty0 819--829, 2011.

\bibitem[Rizopoulos(2012)]{rizopoulos_joint_2012}
D.~Rizopoulos.
\newblock \emph{Joint {{Models}} for {{Longitudinal}} and {{Time}}-to-{{Event
  Data}}: {{With Applications}} in {{R}}}.
\newblock {CRC Press}, 2012.

\bibitem[Schoop et~al.(2008)Schoop, Graf, and
  Schumacher]{schoop_quantifying_2008}
R.~Schoop, E.~Graf, and M.~Schumacher.
\newblock Quantifying the {{Predictive Performance}} of {{Prognostic Models}}
  for {{Censored Survival Data}} with {{Time}}-{{Dependent Covariates}}.
\newblock \emph{Biometrics}, 64\penalty0 (2):\penalty0 603--610, 2008.

\bibitem[{Stan Development Team}(2018)]{rstan2018}
{Stan Development Team}.
\newblock {RStan}: the {R} interface to {Stan}, 2018.
\newblock URL \url{http://mc-stan.org/}.
\newblock R package version 2.17.3.

\bibitem[Steyerberg et~al.(2010)Steyerberg, Vickers, Cook, Gerds, Gonen,
  Obuchowski, Pencina, and Kattan]{steyerberg_assessing_2010}
E.~W. Steyerberg, A.~J. Vickers, N.~R. Cook, T.~Gerds, M.~Gonen, N.~Obuchowski,
  M.~J. Pencina, and M.~W. Kattan.
\newblock Assessing the performance of prediction models: A framework for
  traditional and novel measures.
\newblock \emph{Epidemiology}, 21\penalty0 (1):\penalty0 128--138, 2010.

\bibitem[Sungduk and Albert(2016)]{kim_sungduk2016}
K.~Sungduk and P.~S. Albert.
\newblock A class of joint models for multivariate measurements and a binary
  event.
\newblock \emph{Biometrics}, 72\penalty0 (3):\penalty0 917--925, 2016.

\bibitem[Sutradhar and Ali(1986)]{sutradhar_estimation_1986}
B.~C. Sutradhar and M.~M. Ali.
\newblock Estimation of the parameters of a regression model with a
  multivariate t error variable.
\newblock \emph{Communications in Statistics - Theory and Methods}, 15\penalty0
  (2):\penalty0 429--450, January 1986.

\bibitem[Taylor et~al.(2013)Taylor, Park, Ankerst, Proust-Lima, Williams,
  Kestin, Bae, Pickles, and Sandler]{taylor2013}
J.~M.~G. Taylor, Y.~Park, D.~P. Ankerst, C.~Proust-Lima, S.~Williams,
  L.~Kestin, K.~Bae, T.~Pickles, and H.~Sandler.
\newblock Real-time individual predictions of prostate cancer recurrence using
  joint models.
\newblock \emph{Biometrics}, 69:\penalty0 206--213, 2013.

\bibitem[{van Houwelingen}(2007)]{van_houwelingen_dynamic_2007}
H.~C. {van Houwelingen}.
\newblock Dynamic prediction by landmarking in event history analysis.
\newblock \emph{Scandinavian Journal of Statistics}, 34\penalty0 (1):\penalty0
  70--85, 2007.

\bibitem[Wu(2009)]{wu_mixed_2009}
L.~Wu.
\newblock \emph{Mixed {{Effects Models}} for {{Complex Data}}}.
\newblock {CRC Press}, 2009.

\bibitem[Wulfson and Tsiatis(1997)]{wulfsohn1997}
M.~S. Wulfson and A.~A. Tsiatis.
\newblock A joint model for survival and longitudinal data measured with error.
\newblock \emph{Biometrics}, 53\penalty0 (1):\penalty0 330--339, 1997.

\end{thebibliography}

\end{document}